\documentclass[reprint,aps,prl,superscriptaddress,nobibnotes]{revtex4-1}%
\usepackage{amsmath}
\usepackage{amssymb}
\usepackage{natbib}
\usepackage{graphicx}
\usepackage{xcolor}
\usepackage{amsfonts}%
\providecommand{\U}[1]{\protect\rule{.1in}{.1in}}

\setcitestyle{numbers,square}

\begin{document}

\title{Magnon Accumulation in Chirally Coupled Magnets}
\author{Tao Yu}
\affiliation{Kavli Institute of NanoScience, Delft University of Technology,
	2628 CJ Delft, The Netherlands}
\author{Yu-Xiang Zhang}
\affiliation{Department of Physics and Astronomy, Aarhus University, 8000
	Aarhus C, Denmark}
\author{Sanchar Sharma}
\affiliation{Kavli Institute of NanoScience, Delft University of Technology,
	2628 CJ Delft, The Netherlands}
\author{Yaroslav M. Blanter}
\affiliation{Kavli Institute of NanoScience, Delft University of Technology,
	2628 CJ Delft, The Netherlands}
\author{Gerrit E. W. Bauer}
\affiliation{Institute for Materials Research \& WPI-AIMR \& CSRN, Tohoku
	University, Sendai 980-8577, Japan} 
\affiliation{Kavli Institute of
	NanoScience, Delft University of Technology, 2628 CJ Delft, The Netherlands}
\date{\today }

\begin{abstract}
We report strong chiral coupling between magnons and photons in microwave
waveguides that contain chains of small magnets on special lines. Large magnon
accumulations at one edge of the chain emerge when exciting the magnets by a
phased antenna array. This mechanism holds the promise of new functionalities
in non-linear and quantum magnonics.

\end{abstract}
\maketitle

\textit{Introduction}.---The direct dipolar and exchange interactions between
electron spins in condensed matter create a rich variety of magnetic order
\cite{quantum_magnetism}. The Ruderman-Kittel-Kasuya-Yosida \cite{RKKY}
interaction is mediated by the non-local exchange of mobile electrons.
Magnons, the elementary excitation of the magnetic order, generate the
coupling between nuclear spins \cite{Suhl_Nakamura1,Suhl_Nakamura2}. The range
of these indirect interactions is limited by the coherence length of the
mediator that may be strongly affected by disorder. Photons interact only
weakly with condensed matter but have long coherence lengths
\cite{optical_coherence}, causing interesting and potentially applicable
effects on the magnetic order.

The strong exchange interaction of spins in ferromagnets generates a large
magnetic dipole that couples strongly with photons in high-quality microwave
cavities to create a hybridized quasiparticle---the cavity-magnon polariton
\cite{first_2010,second,third,fourth}. Combining the best features of
high-speed photons and long--lived magnons in low-loss materials such as
yttrium iron garnet (YIG), cavity-magnon polaritons are attractive information
carriers for quantum communication
\cite{first_2010,second,third,fourth,MagnonDarkModes,Lambert_2016,Babak}.
Mediated by the cavity photons with long coherence time, two magnets can be
coupled coherently and tunably over macroscopic distances to create dark and
bright states \cite{MagnonDarkModes,Lambert_2016,Babak}. The counterpart of
coherent coupling---dissipative coupling---between two local spins is
described by non-Hermitian Hamiltonians
\cite{non_hermitian0,non_hermitian1,non_hermitian2,non_hermitian3,non_hermitian4,non_hermitian5,non_hermitian5,non_hermitian6,non_hermitian7,non_hermitian8,non_hermitian9,Book,level_attraction_Canming,level_attraction_Yu}%
, and leads to novel physics such as topological phases
\cite{non_hermitian5,non_hermitian6,non_hermitian7,non_hermitian8,non_hermitian9,Book}
with a non-Hermitian skin effect \cite{non_hermitian6,non_hermitian5}, super-
and sub-radiance
\cite{subradiance1,subradiance2,subradiance3,subradiance4,subradiance5,subradiance6}%
, as well as critical behavior beyond the standard paradigms \cite{D1,D2,D3},
but has not yet been explored in magnetic systems.

\begin{figure}[th]
\vskip 0.25cm {\includegraphics[width=7.8cm]{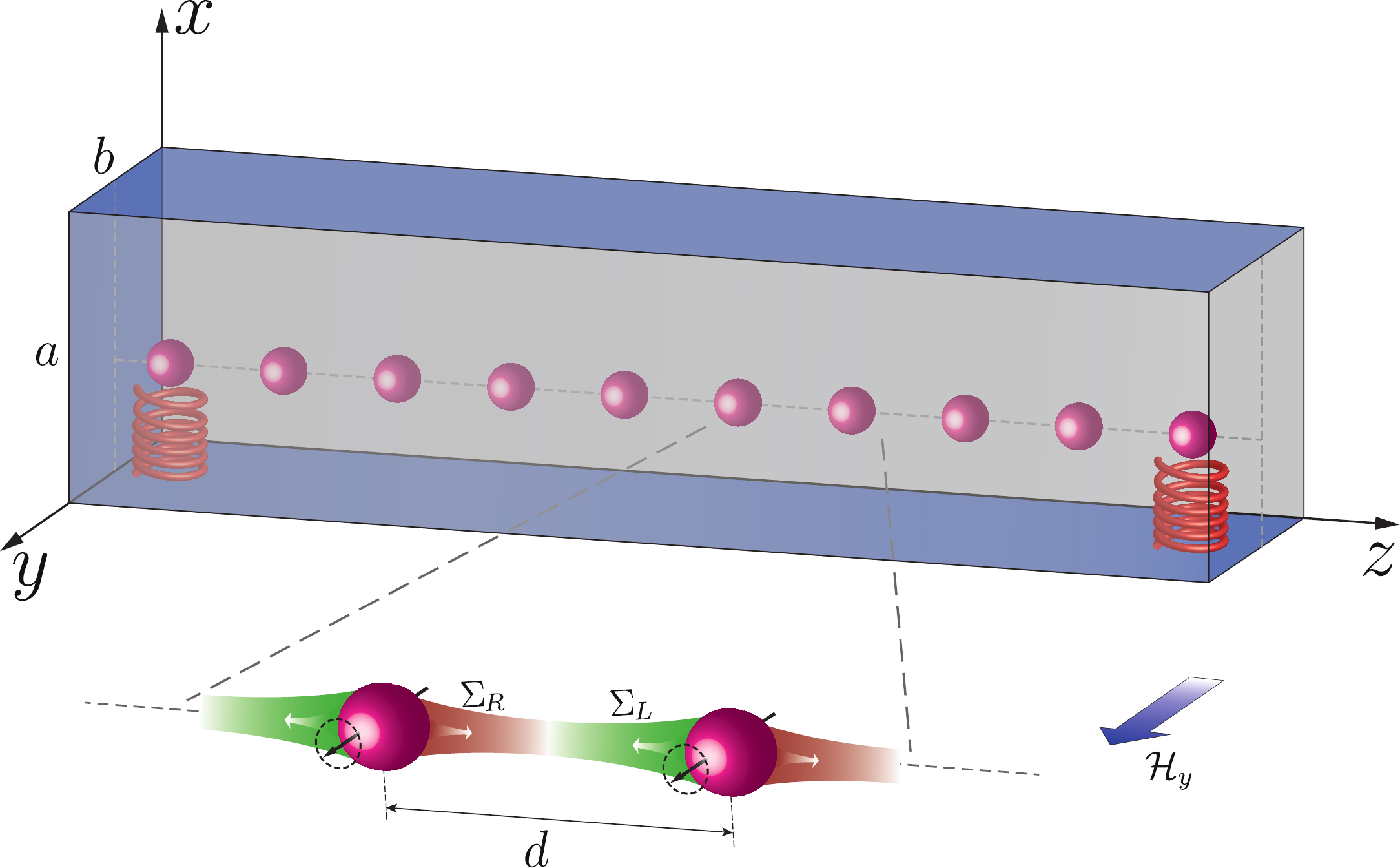}}\caption{Chain of
magnetic spheres with period $d$ in a microwave waveguide. The chain and
waveguide are parallel to the $\hat{z}$-axis and magnetizations are oriented
along $\hat{\mathbf{y}}$ by a magnetic field $H_{y}$. Every magnets interacts
with all other magnets to the right ($\Sigma_{R}$) and left ($\Sigma_{L}$)$.$
Small coils attached to each magnet can excite and detect the local magnon
accumulation (only two are shown).}%
\label{fig:chain}%
\end{figure}

In this Letter, we address the new functionalities that arise when magnetic
particles couple with microwave modes that propagate only in one direction
(chiral coupling). The excited state of a magnet on a line then affects only
the magnets on one side without backaction. Below, we demonstrate that such
chirality can be realized by special positions in a waveguide at which the
precession direction of the photon magnetic field is locked to its wave vector
\cite{Jackson,chiral_optics1,chiral_optics2,chiral_optics4,chiral_optics5,chiral_review}%
. We address the consequences of this chirality in a microwave waveguide as
shown in Fig.~\ref{fig:chain}, which is loaded with a chain of magnets close
to a special line that can individually be addressed by local (coil) antennas
\cite{MagnonDarkModes}. The antenna array allows controlled excitation and detection of
individual magnets as well as collective modes that is not possible by a
global waveguide input and output. We predict that a chiral magnon-photon
coupling in such an array leads to magnon edge states on one side of the
chain. A large magnon amplitude can be generated by relatively weak excitation
power, so our scheme is an\ alternative to parametric pumping
\cite{parametric_pumping1,parametric_pumping2}. We envision that similar
effects occur when a chiral interaction between magnets is mediated by other
quasiparticles such as other magnons \cite{Yu1,Yu2}, conduction electron spins
\cite{non_local} and phonons \cite{Kamra_phonon,Simon_phonon}. This Letter is
accompanied by a longer manuscript \cite{PRB} that exposes the basic theory
and focusses on the microwave scattering through a waveguide by more
configurations including one, two, and many magnets.

\textit{Formalism}.---We consider a waveguide along the $\hat{\mathbf{z}}%
$-direction with a rectangular cross section and model it by Maxwell's
equations with metallic boundary conditions. Even though the predicted effects
obey classical physics, we use a quantum formalism for technical convenience
as well as future research into quantum effects. The total Hamiltonian
$\hat{H}_{\mathrm{em}}+\hat{H}_{\mathrm{m}}+\hat{H}_{\mathrm{int}}$ is of the
Fano-Anderson type \cite{Fano,Mahan,Magnon_radiation}. The photons form a set
of harmonic oscillators:
\begin{equation}
\hat{H}_{\mathrm{em}}=\int\hbar\Omega_{k}\hat{p}_{k}^{\dagger}\hat{p}_{k}dk,
\end{equation}
where $\hat{p}_{k}$ annihilates a photon with frequency $\Omega_{k}$ and
momentum $k$ of the lowest transverse $\mathrm{TE}_{10}$ mode that is
polarized along and uniformly distributed over the $\hat{\mathbf{y}}%
$-direction and with standing wavelength $2a$ in the $\hat{\mathbf{x}}%
$-direction (see Fig.~\ref{fig:chain} with $a>b$). The dispersion{\textit{\ }%
}$\Omega_{k}^{2}/c^{2}=k^{2}+(m/a)^{2}+(n/b)^{2}\ $and from now on{\textit{\ }%
}$\{m,n\}=\{1,0\}$.

The waveguide is loaded with $N$ identical YIG spheres with gyromagnetic ratio
$-\gamma$, saturation magnetization $M_{s}$, and volume $V_{s}$ at
$\mathbf{r}_{j}=(\pmb{\rho},(j-1)d)$ with $j\in\{1,2,\cdots,N\}$, where
$\pmb{\rho}=(x,y)$ and $d$ is the (equidistant) spacing between the spheres.
The sub-mm spheres are much smaller than the photon wavelength of
$\mathcal{O}\left(  \mathrm{cm}\right)  $, so they can be treated as point
particles. The static magnetic field $H_{y}$ in the $\hat{\mathbf{y}}%
$-direction in Fig.~\ref{fig:chain} is sufficiently strong to saturate the
magnetization. The waveguide photons couple to the anti-clockwise uniform
magnetization precession around the magnetic field (Kittel mode). In second
quantization
\begin{equation}
M_{j,z}-iM_{j,x}=\sqrt{{2\hbar\gamma M_{s}}/{V_{s}}}\hat{m}_{j},
\label{Quant:Mag}%
\end{equation}
where $M_{j,\delta}$ is the $\delta$-th component of the magnetization
amplitude in the $j$-th magnet and $\hat{m}_{j}$ is a bosonic annihilation
operator. The dynamics is governed by the linearized Landau-Lifshitz equation
(damping to be included below), which after substituting Eq.~(\ref{Quant:Mag})
reduces to a harmonic oscillator for each magnet
\begin{equation}
\hat{H}_{\mathrm{m}}=\hbar\omega_{m}\sum_{j=1}^{N}\hat{m}_{j}^{\dagger}\hat
{m}_{j},
\end{equation}
where $\omega_{m}=\mu_{0}\gamma H_{y}$ is the Larmor precession frequency and
$\mu_{0}$ is the vacuum permeability.

The photons and magnons are coupled by the Zeeman interaction
\begin{equation}
\hat{H}_{\mathrm{int}}=\sum_{j}\int\frac{dk}{\sqrt{2\pi}}\left[  \hbar
g_{j}(k)\hat{p}_{k}\hat{m}_{j}^{\dagger}+\mathrm{h.c.}\right]  .
\label{Ham:Int}%
\end{equation}
The coupling constant $g_{j}(k)=\tilde{g}(k)e^{ik(j-1)d}$, where
\begin{equation}
\tilde{g}(k)=-\mu_{0}\sqrt{\frac{\gamma M_{s}V_{s}}{2\hbar}}\mathcal{H}%
_{k,-}\left(  \pmb{\rho}\right)  ,
\end{equation}
with $\mathcal{H}_{k,-}=\mathcal{H}_{k,z}-i\mathcal{H}_{k,x}$ being the microwave
transverse magnetic field.

The magnons interact resonantly with photons with wave numbers near
$k_{0}=\sqrt{\omega_{m}^{2}/c^{2}-\pi^{2}/a^{2}}$. The magnetic field of
$\mathrm{TE}$-photons is polarization-momentum locked, i.e. $\mathcal{H}%
_{k,-}$ depends on the sign of $k$
\cite{Jackson,chiral_optics1,chiral_optics2,chiral_optics4,chiral_optics5,chiral_review}%
. A magnet, e.g., kept at a position where $\mathcal{H}_{-k_{0},-}$ is zero
but $\mathcal{H}_{k_{0},-}$ is finite, can radiate only into the
$+\hat{\mathbf{z}}$-direction. For the $\mathrm{TE}_{10}$ mode this occurs for
arbitrary $y$ and
\begin{equation}
\cot\left(  {\pi x}/{a}\right)  =-\sqrt{k_{0}a}{\pi}. \label{ChiralPoint}%
\end{equation}

The effective coupling between spheres can be modelled by integrating out the
photon fields (for details see Ref.~\cite{PRB}) in terms of the equation of
motion for the vector of magnetizations $\hat{\mathcal{M}}=\left(  \hat{m}%
_{1},\dots,\hat{m}_{N}\right)  ^{T}$
\begin{equation}
{d\hat{\mathcal{M}}}/{dt}=-i\tilde{H}_{\mathrm{eff}}\hat{\mathcal{M}}%
-\hat{\mathcal{T}}-\sqrt{\alpha_{G}\omega_{m}}\hat{\mathcal{N}}.
\label{EOM:Mags}%
\end{equation}
$\alpha_{G}\ $is the Gilbert damping constant and $\hat{\mathcal{T}}%
=\hat{\mathcal{T}}_{w}+\hat{\mathcal{T}}_{l}$ is the external torque by the
wave guide photons
\begin{equation}
\hat{\mathcal{T}}_{w}=i\int\frac{dk}{\sqrt{2\pi}}\tilde{g}(k)\hat
{p}_{k,\mathrm{in}}e^{-i\Omega_{k}t}\left(  e^{ikd},\cdots,e^{ikNd}\right)
^{T}, \label{waveguide_input}%
\end{equation}
and the local antennas $\hat{\mathcal{T}}_{l}=(\hat{P}_{1}(t),\cdots,\hat
{P}_{N}(t))^{T}$. $\hat{\mathcal{N}}=\left(  \hat{n}_{1},\cdots,\hat{n}%
_{N}\right)  ^{T}$ is the thermal noise in the magnetic system. We model
$\hat{n}_{j}$ as white noise sources satisfying $\langle{\hat{n}_{j}}%
\rangle=\langle{\hat{n}_{j}\hat{n}_{i\neq j}}\rangle=0$, $\langle{\hat{n}%
_{j}^{\dagger}(t)\hat{n}_{j}(t^{\prime})}\rangle=n_{m}\delta(t-t^{\prime})$,
$\langle{\hat{n}_{j}(t)\hat{n}_{j}^{\dagger}(t^{\prime})}\rangle
=(n_{m}+1)\delta(t-t^{\prime})$, where $n_{m}=1/\left\{  \exp\left[
{\hbar\omega_{m}}/({k_{B}T})\right]  -1\right\}  $ is the thermal occupation
of magnons at (constant) bath temperatures $T>\hbar\omega_{m}/k_{B}$
\cite{input_output1,input_output2}.

In the non-Hermitian matrix $\tilde{H}_{\mathrm{eff}}=\tilde{\omega}+\Sigma$,
$\tilde{\omega}_{jl}\equiv\tilde{\omega}_{m}\delta_{jl}=\omega_{m}\left(
1-i\alpha_{G}\right)  \delta_{jl}$, and the photon-mediated self-energy
\begin{equation}
\Sigma_{jl}=-i%
\begin{cases}
(\Gamma_{L}+\Gamma_{R})/2, & j=l\\
\Gamma_{R}e^{ik_{0}(j-l)d}, & j>l\\
\Gamma_{L}e^{ik_{0}(l-j)d}, & j<l
\end{cases}
, \label{Def:Sigma}%
\end{equation}
where $\Gamma_{R}=\tilde{g}^{2}(k_{0})/v(k_{0}),\Gamma_{L}=\tilde{g}%
^{2}(-k_{0})/v(k_{0})$ and $v(k)=|k|c^{2}/\omega_{m}$ is the photon group
velocity. The self-energy contributes to the dissipative and long-range
coupling between any two magnets. The chiral coupling appears when $\Gamma
_{L}\neq\Gamma_{R}$. The direct coupling between any two magnets does not
depend on distance because we may safely disregard retardation and assume
sufficiently high quality of the waveguide and magnets.

\textit{Hamiltonian.}---The Hamiltonian $\tilde{H}_{\mathrm{eff}}$ is
non-Hermitian, but can be diagonalized by introducing left and right
eigenvectors \cite{Book}. The right eigenvectors of $\Sigma$, say
$\{\psi_{\zeta}\}$ with corresponding eigenvalues $\{\gamma_{\zeta}\}$,
satisfy $\left(  \gamma_{\zeta}-\Sigma\right)  \psi_{\zeta}=0$ for a
delocalized mode with label $\zeta\in\{1,\dots,N\}$. Here $\mathrm{Re} \left[
\gamma_{\zeta}\right]  $ is the resonance frequency and $\mathrm{Im}\left[
\gamma_{\zeta}\right]  $ the reciprocal lifetime. $\{\phi_{\zeta}\}\neq
\{\psi_{\zeta}\}$ are the eigenvectors of $\Sigma^{\dagger}$ with eigenvalues
$\{\gamma_{\zeta}^{\ast}\}$. In the absence of degeneracies in $\{\gamma
_{\zeta}\}$ the (normalized) modes are \textquotedblleft
bi-orthonormal\textquotedblright, i.e. $\phi_{\zeta}^{\dagger}\psi
_{\zeta^{\prime}}=\delta_{\zeta\zeta^{\prime}}$ \cite{subradiance1}. With
$\Sigma^{\dagger}=\mathcal{P}\Sigma^{\ast}\mathcal{P}$ where $\mathcal{P}%
_{ij}=\delta_{i+j,N}$ inverts the order of magnets $1\leftrightarrow N$,
$2\leftrightarrow N-1,$ $\cdots$, we arrive at $\phi_{\zeta}=\mathcal{P}%
\psi_{\zeta}^{\ast}$ \cite{PRB}.

$\tilde{H}_{\mathrm{eff}}$ consists of a Hermitian $\tilde{H}_{h}=(\tilde
{H}_{\mathrm{eff}}+\tilde{H}_{\mathrm{eff}}^{\dagger})/2$ and non-Hermitian
part $\tilde{H}_{nh}=(\tilde{H}_{\mathrm{eff}}-\tilde{H}_{\mathrm{eff}%
}^{\dagger})/2$. $\hat{\mathcal{M}}$ can be expanded into generalized Bloch
states $\hat{\Psi}_{\kappa}=\sum_{j=1}^{N}e^{i{\kappa}z_{j}}\hat{m}_{j}%
/\sqrt{N}$ with $z_{j}=(j-1)d$ and complex \textquotedblleft crystal
momentum\textquotedblright\ $\kappa$. Two Bloch states $\hat{\Psi}_{k_{0}}$
and $\hat{\Psi}_{-k_{0}}$ diagonalize $\tilde{H}_{nh}$ (recall $k_{0}%
=\sqrt{\omega_{m}^{2}/c^{2}-\pi^{2}/a^{2}}$)
\begin{equation}%
\begin{pmatrix}
\nu+\frac{i}{2}\Gamma_{L}N & \frac{i}{2}\Gamma_{R}\frac{1-e^{2ik_{0}Nd}%
}{1-e^{2ik_{0}d}}\\
\frac{i}{2}\Gamma_{L}\frac{1-e^{-2ik_{0}Nd}}{1-e^{-2ik_{0}d}} & \nu+\frac
{i}{2}\Gamma_{R}N
\end{pmatrix}%
\begin{pmatrix}
\hat{\Psi}_{k_{0}}\\
\hat{\Psi}_{-k_{0}}%
\end{pmatrix}
=0.
\end{equation}
The sum of the eigenvalues $\nu_{+}+\nu_{-}=-iN(\Gamma_{L}+\Gamma_{R})/2$ is
the total radiative decay rate, which scales with the number of magnets. These
two states are called \textquotedblleft superradiant\textquotedblright\ or
\textquotedblleft bright\textquotedblright, while the remaining $(N-2)$ states
are \textquotedblleft subradiant\textquotedblright\ or \textquotedblleft
dark\textquotedblright\ with initially infinite radiative lifetime. The
coherent coupling by $\tilde{H}_{h}$ mixes all states, but subradiant states
with enhanced lifetimes persist, as we show below by a combined analytic and
numerical treatment (see also Ref.~\cite{PRB}).

The ansatz of extended Bloch states $\hat{\Psi}_{\kappa}$ leads to the closed
expression for the homogeneous Schr\"{o}dinger equation \cite{subradiance2}
\begin{equation}
d\hat{\Psi}_{\kappa}/dt=-i\omega_{\kappa}\hat{\Psi}_{\kappa}-\Gamma
_{L}g_{\kappa}\hat{\Psi}_{k_{0}}+\Gamma_{R}h_{\kappa}\hat{\Psi}_{-k_{0}%
},\label{EOM_Bloch_states}%
\end{equation}
in which
\begin{equation}
\omega_{\kappa}\equiv-i\frac{\Gamma_{R}}{2}\frac{1+e^{i({\kappa}+k_{0})d}%
}{1-e^{i({\kappa}+k_{0})d}}+i\frac{\Gamma_{L}}{2}\frac{1+e^{i({\kappa}%
-k_{0})d}}{1-e^{i({\kappa}-k_{0})d}},\label{omk}%
\end{equation}
with $g_{\kappa}=1/[1-e^{i({\kappa}-k_{0})d}]$ and $h_{\kappa}={e^{i({\kappa
}+k_{0})Nd}}/[1-e^{i({\kappa}+k_{0})d}].$ In an infinite chain (or a closed
ring) $\hat{\Psi}_{\kappa}$ would be a solution. The boundary conditions of
the finite system can be fulfilled by the superposition of two states with
momenta $\kappa$ and $\kappa^{\prime}$ at the same frequency $\omega_{\kappa
}=\omega_{\kappa^{\prime}}$. The additional terms appearing in
Eq.~(\ref{EOM_Bloch_states}) are cancelled by enforcing
\begin{equation}
g_{\kappa}h_{\kappa^{\prime}}=g_{\kappa^{\prime}}h_{\kappa}%
,\label{coupled_equation}%
\end{equation}
leading to eigenstates $\hat{\alpha}_{\zeta}=\sum_{j}\phi_{\zeta,j}^{\ast}%
\hat{m}_{j}\propto(g_{\kappa}\hat{\Psi}_{\kappa^{\prime}}-g_{\kappa^{\prime}%
}\hat{\Psi}_{\kappa}).$ The wave functions and spectra then read
\begin{equation}
\psi_{\zeta,j}\propto g_{\kappa^{\prime}}e^{i\kappa z_{N-j}}-g_{\kappa
}e^{i\kappa^{\prime}z_{N-j}};\;\gamma_{\zeta}=\omega_{\kappa}.\label{solution}%
\end{equation}
Only when the system is inversion symmetric ($\Gamma_{L}=\Gamma_{R}$), the
solutions reduce to standing waves with $\operatorname{Re}\kappa^{\prime
}=-\operatorname{Re}\kappa$.

$\omega_{\kappa}$ diverges at $\kappa=\pm k_{0}$. On the other hand, the
radiative damping $\sim\mathrm{\operatorname{Im}}\omega_{\kappa}$ is minimized
for say ${\kappa}={\kappa}_{\ast}.$ Neither $\kappa=\pm k_{0}$ nor
$\kappa_{\ast}$ solve Eq.~(\ref{coupled_equation}), but these states reflect
the \textquotedblleft superradiance\textquotedblright\ and \textquotedblleft
subradiance" well-known in quantum optics
\cite{subradiance1,subradiance2,subradiance3,subradiance4,subradiance5,subradiance6}%
. The former corresponds to the edge states of $\tilde{H}_{\mathrm{eff}}%
\ $with enhanced magnon amplitudes and damping, while the latter are weakly
coupled delocalized standing waves, as demonstrated in the following.

The wave numbers ${\kappa}_{\ast}$ of the extremal points
$\mathrm{\operatorname{Im}}\omega_{{\kappa}_{\ast}}$ obey
\begin{align}
{\kappa}_{\ast}d &  =\arcsin\frac{\Gamma_{R}-\Gamma_{L}}{\sqrt{\Gamma_{R}%
^{2}+\Gamma_{L}^{2}-2\Gamma_{R}\Gamma_{L}\cos(2k_{0}d)}}\nonumber\\
&  -\arctan\frac{\Gamma_{R}-\Gamma_{L}}{(\Gamma_{R}+\Gamma_{L})\tan(k_{0}d)}.
\end{align}
$\arcsin x$ is a two-valued function in the first Brillouin zone $[-\pi
/d,\pi/d]$ and we have two extremal points. The two solutions close to each
extremum ${\kappa}_{\pm}={\kappa}_{\ast}\pm\delta$ label degenerate states
that solve $g_{{\kappa}_{+}}h_{{\kappa}_{-}}=g_{{\kappa}_{-}}h_{{\kappa}_{+}}%
$. For small $\delta$,
\begin{equation}
\delta_{\zeta}=\frac{\zeta\pi}{Nd}\left[  1-\frac{i}{N}\frac{\sin\left(
k_{0}d\right)  }{\cos\left(  {\kappa}_{\ast}d\right)  -\cos\left(
k_{0}d\right)  }\right]  ,
\end{equation}
where $\zeta\in%
\mathbb{N}
$. With Eq.~(\ref{solution}), the wave function and dispersion of these
subradiant states read
\begin{align}
\psi_{\zeta,j} &  \approx-2i\frac{e^{i\kappa_{\ast}z_{N-j}}}{1-e^{i(\kappa
_{\ast}-k_{0})d}}\sin(\delta_{\zeta}z_{N-j}),\nonumber\\
\omega_{\zeta} &  =\omega_{{\kappa}_{\ast}}+\frac{\sin(k_{0}d)}{\cos({\kappa
}_{\ast}d)-\cos(k_{0}d)}\frac{\Gamma_{R}(\delta_{\zeta}d)^{2}/2}{1-\cos
[(k_{0}+{\kappa}_{\ast})d]},\label{scaling}%
\end{align}
where $\delta_{\zeta}\propto\zeta/(Nd)$. These solutions are nearly standing
waves with long radiative lifetimes and are only weakly affected by chirality.

We have to numerically calculate the solutions for $\kappa$ close to $\pm
k_{0}$, i.e. $\kappa=k_{0}+\eta$ and $\kappa^{\prime}=-k_{0}+\eta^{\prime}$ in
which $\eta$ and $\eta^{\prime}$ are small complex numbers. $\operatorname{Im}%
\eta$ and $\operatorname{Im}\eta^{\prime}$ govern the decay of the states at
the two edges. With chirality, only one of them is important, which causes a
concentration at one edge of the chain.

As an example, we consider a rectangular waveguide with dimensions $a=1.6$~cm
and $b=0.6$~cm, and $20$ YIG magnetic spheres with radius $r_{c}=0.6$~mm and
$\alpha_{G}=5\times10^{-5}$ \cite{Magnon_radiation}. $\omega_{m}/(2\pi
)=\sqrt{3}c/\left(  2a\right)  =16.2$~GHz is tuned to correspond to the photon
momentum $k_{0}=\sqrt{2}\pi/a$ of the lowest TE$_{10}$ mode. By varying the
position and size of the magnets we may tune the magnon-photon interaction
Eq.~(\ref{Def:Sigma}), here $\Gamma_{R,L}/(2\pi)\in(0,20)$~MHz, while
$\alpha_{G}\omega_{m}/(2\pi)\sim1$~MHz and chiralities $0<\Gamma_{L}%
/\Gamma_{R}<\infty$. $k_{0}d=\pi/5$, corresponds to an intermagnet spacing
$d=a/(5\sqrt{2})\approx0.2$~cm and $Nd\approx4$~cm. The predicted features do
not depend strongly on the chain lengths and are still prominent for a small
number of spheres \cite{PRB}.

\begin{figure}[th]
\begin{center}
{\includegraphics[width=4.41cm]{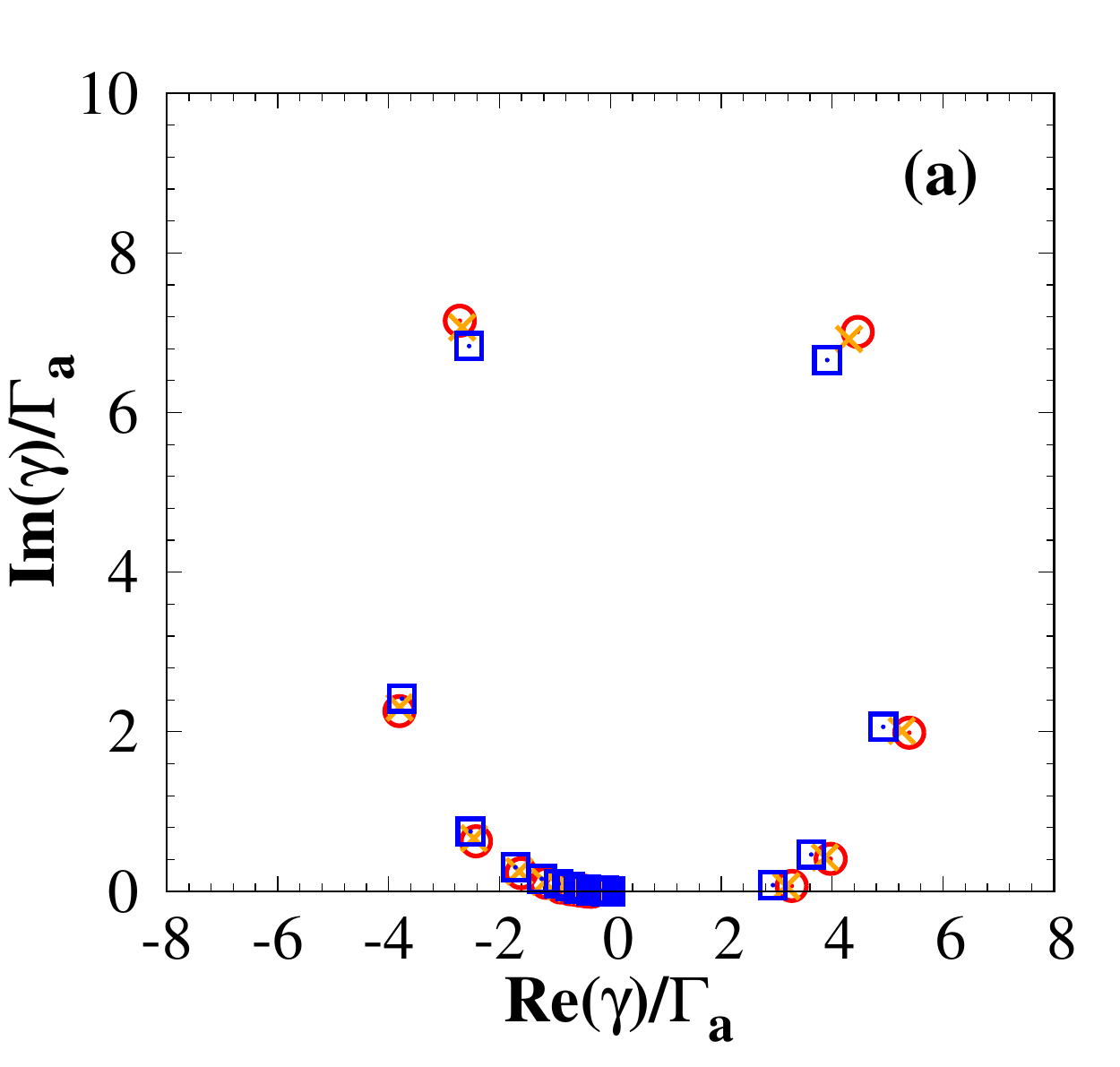}} \hspace{-0.4cm}
{\includegraphics[width=4.41cm]{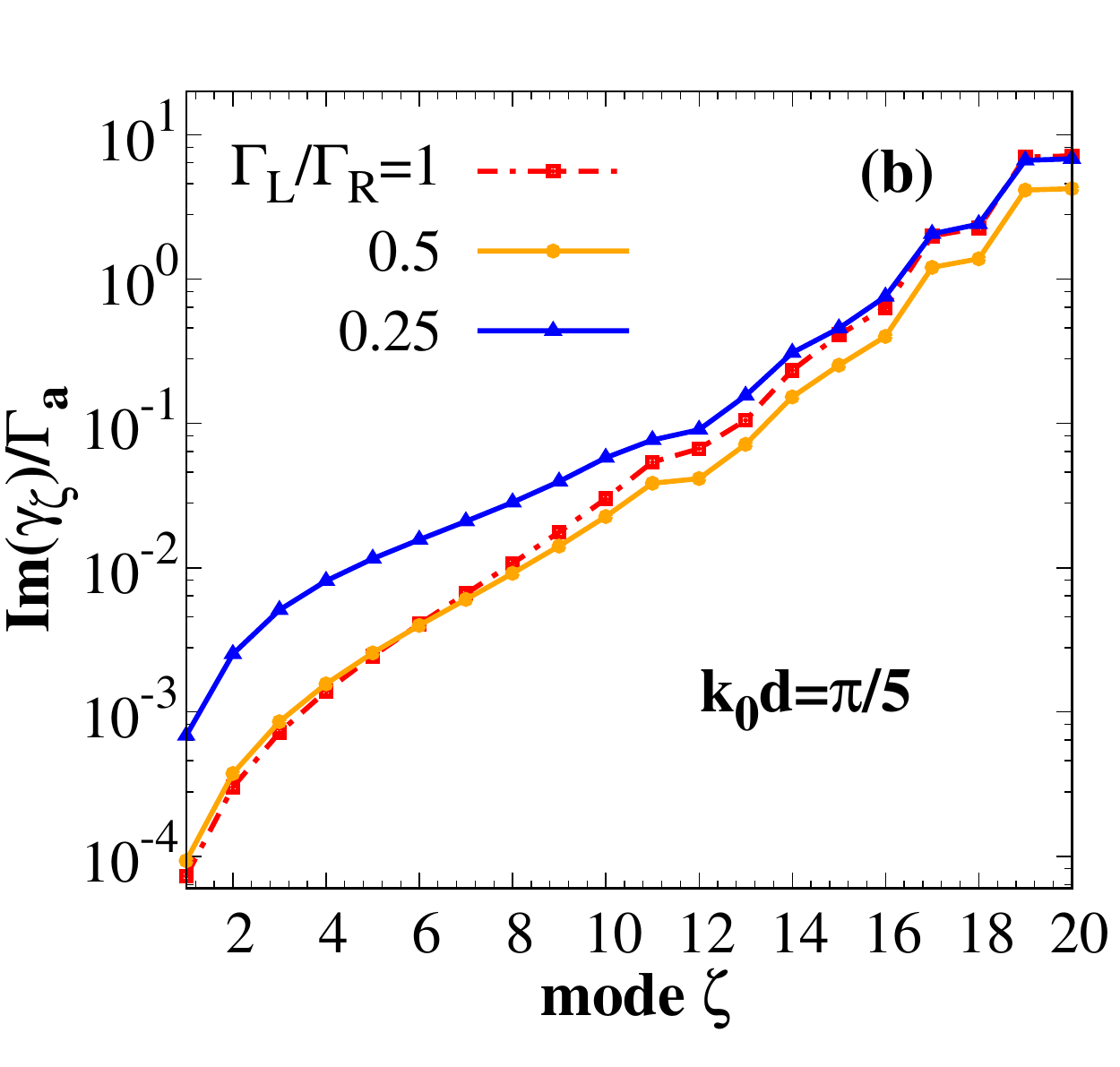}}
{\includegraphics[width=4.41cm]{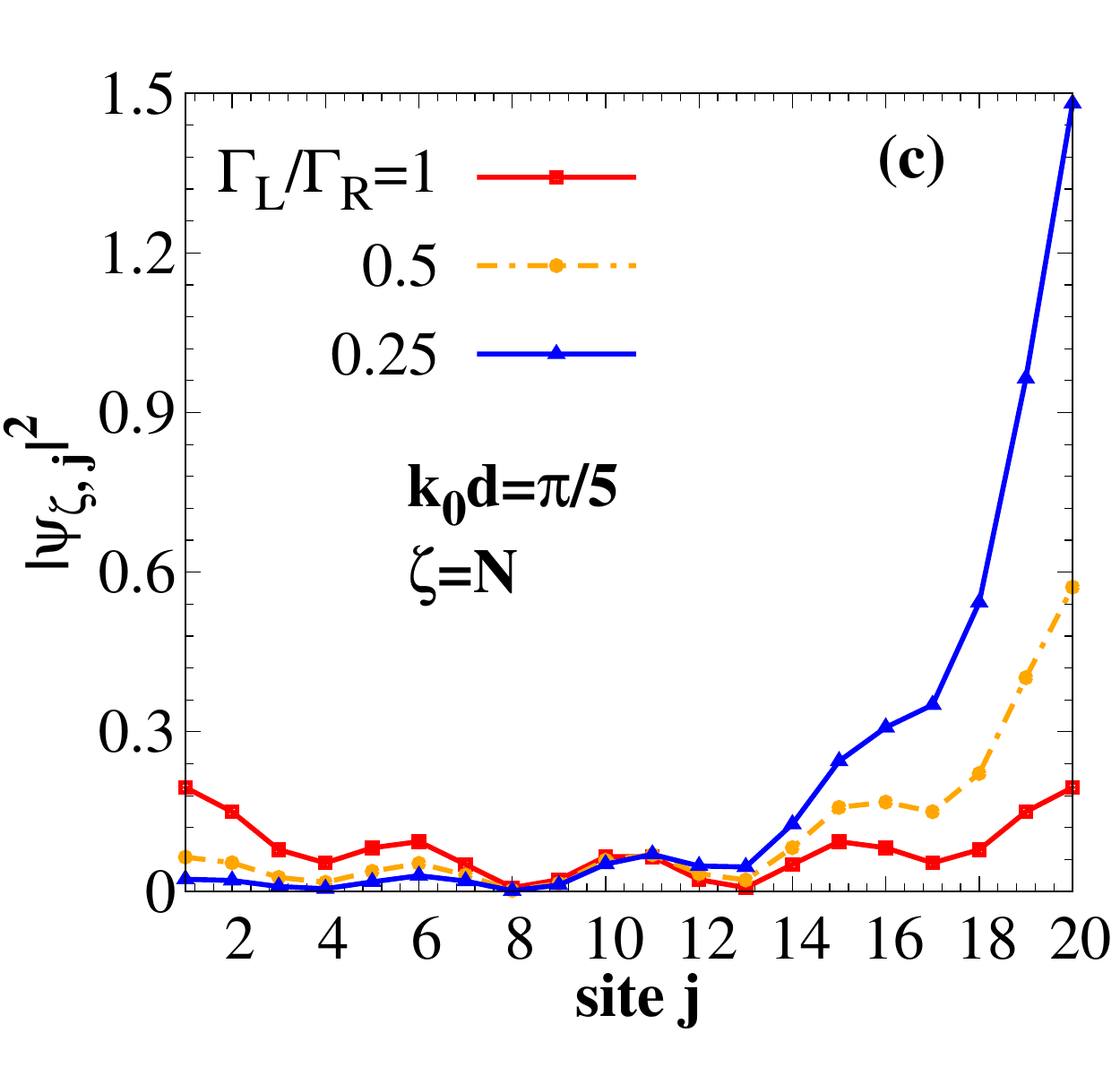}} \hspace{-0.4cm}
{\includegraphics[width=4.41cm]{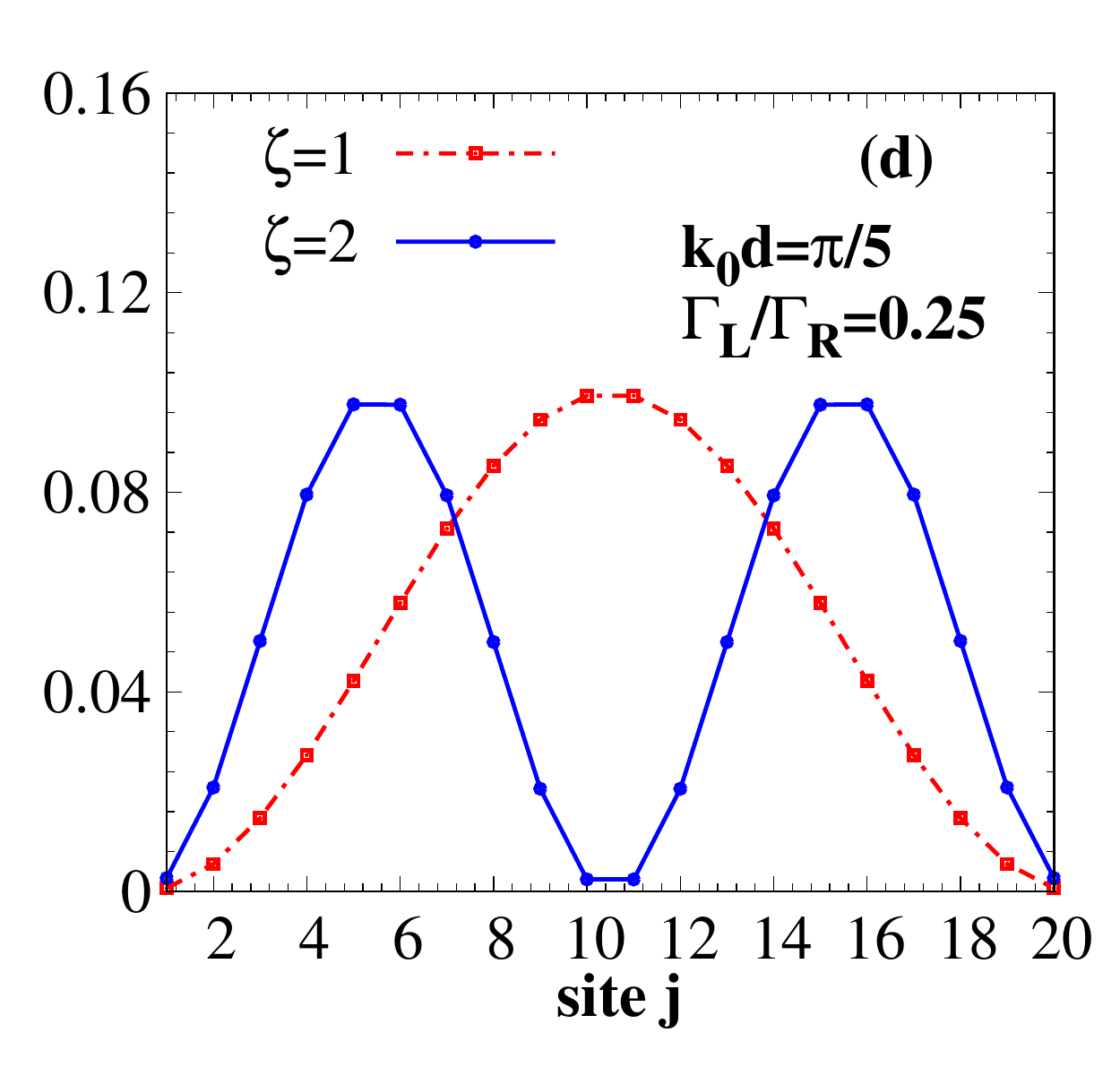}}
\end{center}
\caption{(Color online) Energy spectra and wave functions of the magnet chain
of $20$ magnetic spheres. (a) Real and imaginary components of the eigen
energies $\gamma=(\nu-\tilde{\omega}_{m})$ scaled by $\Gamma_{a}=(\Gamma
_{L}+\Gamma_{R})/2$. Red circles, orange crosses and blue squares encode the
chiralities $\Gamma_{L}/\Gamma_{R}=1$, 0.5 and 0.25, respectively. (b) All
$20$ eigenstates sorted by increasing decay rates. (c) Magnon intensity
distribution of the most short-lived state with $\zeta=20$ in (b). (d) Magnon
intensity distribution for the longest living states with $\zeta=1,2$ in (b).
}%
\label{wavefunction}%
\end{figure}

\textit{Magnon accumulation.}---Figure~\ref{wavefunction} is a plot of the
energy spectra and magnon accumulation (squared wave functions).
Fig.~\ref{wavefunction}(a) shows that the real and imaginary components of the
eigen-energy $\gamma_{\zeta}$, scaled by $\Gamma_{a}=(\Gamma_{L}+\Gamma
_{R})/2$, are approximately distributed on an ellipse in the complex plane
that depends only weakly on the chirality. The solutions with long lifetimes
are clustered around the frequencies $\omega_{\kappa_{\ast}}.$ It is negative
in Fig.~\ref{wavefunction}(a) but depends strongly on $k_{0}.$ Modes with
$\mathrm{\operatorname{Im}}\gamma>\Gamma_{a}$ ($<\Gamma_{a}$) are
super-radiant (subradiant) with radiative lifetime shorter (longer) than that
of an isolated magnet. The decay rates of all eigenstates are sorted and
plotted with integer labels $\zeta\in\{1,2,\cdots,20\}$ in
Fig.~\ref{wavefunction}(b). Here, the typical radiative lifetime of the most
superradiant state $\left(  \zeta=20\right)  $ is $20\sim70$~MHz for the three chiralities.

The magnon accumulation $|\psi_{\zeta,j}|^{2}$ of the most short-lived state
($\zeta=20$ in Fig.~\ref{wavefunction}(b)) is plotted in
Fig.~\ref{wavefunction}(c). When $\Gamma_{R}=\Gamma_{L}$, the state is
symmetrically localized close to both edges (red solid curve), but with
increasing chirality, the distribution becomes asymmetrically skewed to one
boundary. When $\Gamma_{R}<\Gamma_{L}$ ($\Gamma_{R}>\Gamma_{L}$), the boundary
state is localized at the left (right) boundary of the chain. The enhanced
dynamics associated to large magnon numbers causes superradiance. The most
subradiant states, on the other hand, have magnon accumulations $\sim
\left\vert \sin\left(  \zeta\pi j/N\right)  \right\vert ^{2}$, with small
amplitudes at the two boundaries, as shown in Fig.~\ref{wavefunction}(d), and
are only weakly affected by chirality. A weak higher harmonic reflects the
bare photon wave length $\sim2\pi/k_{0}$.

We can now expand the magnetization $\hat{\mathcal{M}}(t)=\sum_{\zeta=1}%
^{N}\hat{\alpha}_{\zeta}(t)\psi_{\zeta}$ into the above eigenstates with
coefficients $\hat{\alpha}_{\zeta}(t)=\phi_{\zeta}^{\dagger}\hat{\mathcal{M}%
}(t)$. For the local input vector at common frequency $\omega_{\mathrm{in}}$,
$\left\langle \hat{\mathcal{T}}_{l}(t)\right\rangle =ie^{-i\omega
_{\mathrm{in}}t}\left(  P_{1},P_{2},\cdots,P_{N}\right)  $ and waveguide
photon feed $\left\langle \hat{\mathcal{T}}_{w}\right\rangle =0$ (we discuss
the case with $\left\langle \hat{\mathcal{T}}_{w}\right\rangle \neq0$ and
$\left\langle \hat{\mathcal{T}}_{l}\right\rangle =0$ in Ref.~\cite{PRB}), the
coherent magnetization amplitude%
\begin{equation}
\left\langle \hat{\mathcal{M}}(t)\right\rangle =-i\sum_{\zeta}\frac
{(\mathcal{P}\psi_{\zeta})^{T}\left\langle \hat{\mathcal{T}}_{l}%
(t)\right\rangle }{\omega_{\mathrm{in}}-\tilde{\omega}_{m}-\gamma_{\zeta}}%
\psi_{\zeta}.\label{M_summation}%
\end{equation}
We are looking for a large magnon accumulation at one edge of the chain due to
the chirality. Since $(\mathcal{P}\psi_{\zeta})^{T}=(\psi_{\zeta,N}%
,\psi_{\zeta,N-1},\cdots,\psi_{1})$ oscillates between spheres with fixed
phase, the vector product $(\mathcal{P}\psi_{\zeta})^{T}\left\langle
\hat{\mathcal{T}}_{l}(t)\right\rangle $ can be large for a localized edge
state ${\zeta_{\ast}}$ on the right when the input from the local antennas
matches its phase and frequency. To match the phases of the edge states, we
consider local power injection of the form $\left\langle \hat{\mathcal{T}}%
_{l}(t)\right\rangle =iP(1,e^{i\phi},\cdots,e^{i(N-1)\phi})\exp[-i(\omega
_{m}+\mathrm{\operatorname{Re}}\gamma_{\zeta_{\ast}})t]$, in which the optimal
phase depends on the number of magnets but $\phi$ $\rightarrow k_{0}d$ for
sufficiently long chains.

\begin{figure}[th]
{\includegraphics[width=4.41cm]{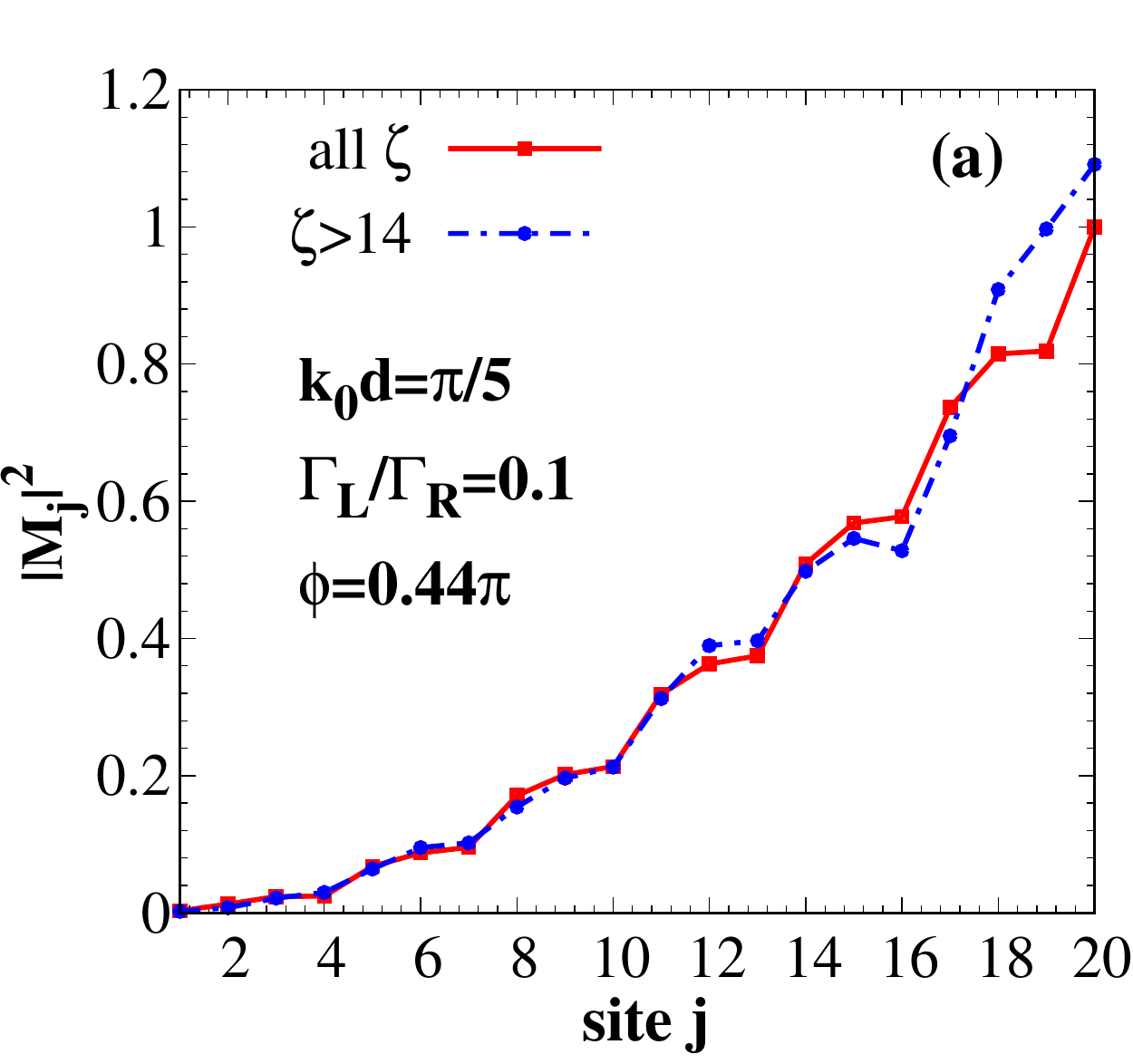}} \hspace{-0.4cm}
{\includegraphics[width=4.41cm]{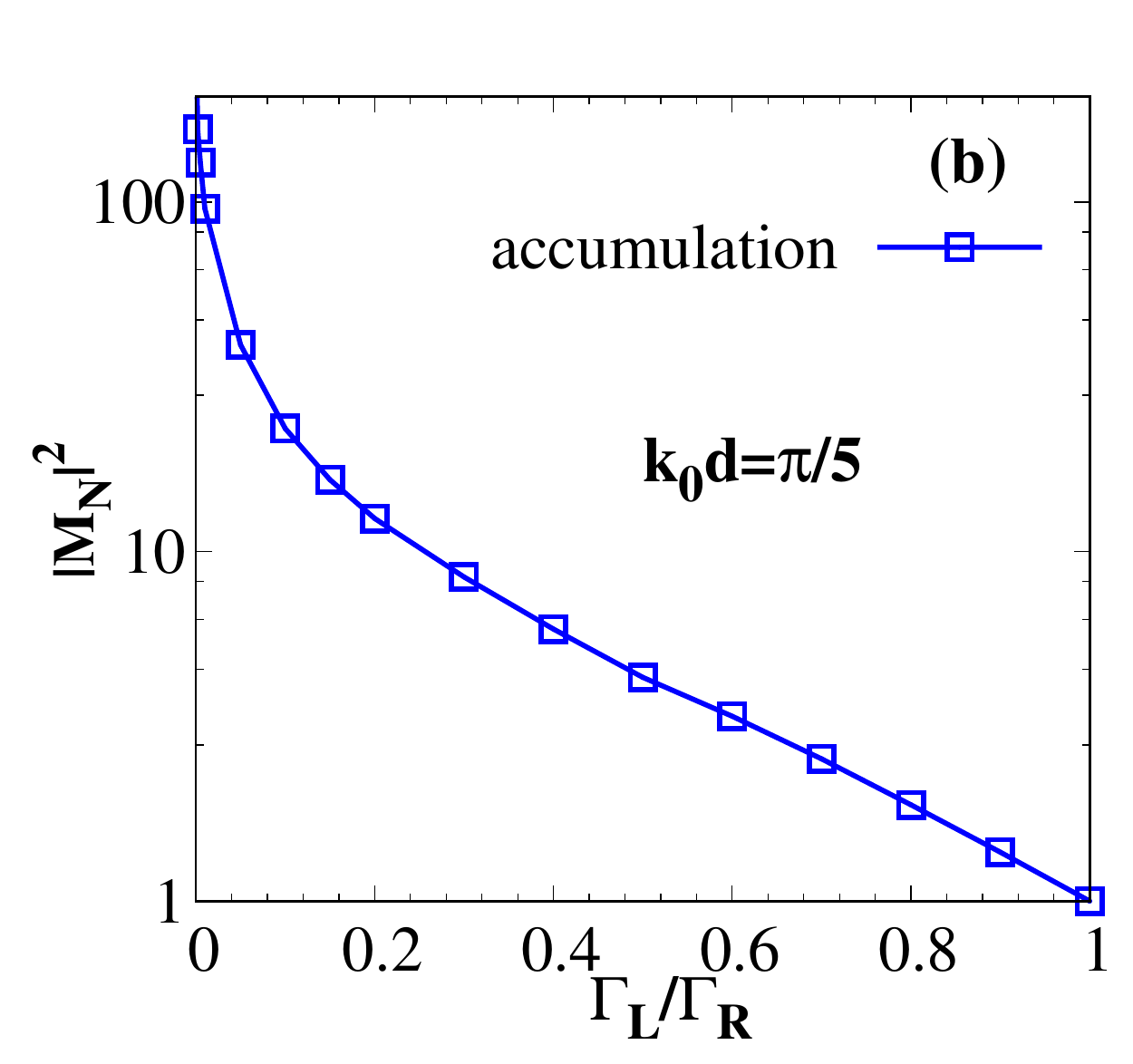}} \caption{(Color online)
Magnon accumulation excited by a phased antenna array. (a), The accumulation
(squared magnetization amplitude) distribution over the magnets (red cureve)
is normalized by the largest value at the edge. The blue curve excludes the
contribtuion of subradiant states. (b), The magnon accumulation at the right
edge as a function of chirality, normalized by the one for $\Gamma_{L}%
/\Gamma_{R}=1$.}%
\label{accumulation}%
\end{figure}

Figure~\ref{accumulation}(a) shows that switching on the local antennas for
$\Gamma_{L}/\Gamma_{R}=0.1$ and phase $\phi$ optimally chosen to be
$\sim0.44\pi$ leads to an enhanced accumulation on the right side. This choice
of $\phi$ is out of phase with the subradiant states that are therefore hardly
excited (see the blue curve in Fig.~\ref{accumulation}(a)).
Fig.~\ref{accumulation}(b) is the accumulation on the right-most sphere as a
function of chirality, which is enhanced more than 100-fold by tuning the
chirality $\Gamma_{L}/\Gamma_{R}\rightarrow0.$ In this limit the frequencies
become degenerate, but individual modes can still be accessed by the phased array.

\textit{Conclusions}.---In conclusion, the interaction between magnons and
photons can be chiral and tunable by strategically positioning small magnets
in a waveguide. We predict a strong imbalance of magnon populations in a chain
of magnets in which the dissipative and long-range nature of the coupling can
strongly enhance the magnon intensity at the edges. The coherent amplitudes of
magnons at one edge of the sample can be much higher than those excited by
conventional ferromagnetic resonance. This allows studying non-linear effects
at low input power. On the other hand, the magnon number of the magnets in the
center of the chain are only weakly affected.

Our formalism can be extended into the quantum regime of magnons, in which we
can profit from the insights of the field of quantum optics, which studies
collective coupling with atomic emitters
\cite{subradiance1,subradiance2,subradiance3,subradiance4,subradiance5}. The
strong coupling between magnons and photons in a microwave waveguide
\cite{Magnon_radiation} opens the new perspective of magnonic quantum emitters
\cite{magnonics4}, which might help circumventing the harsh experimental
environment such as extremely low temperature and fine control required for
cold atom system. We also find analogies with chiral optics, in which the
coupling between light and emitters depends on the propagation of light and
polarization of the local emitters \cite{chiral_review}. The chiral coupling
between emitters is promising in achieving quantum state transfer between
qubits via the magnonic chiral quantum channel introduced here.

\vskip0.25cm \begin{acknowledgments}
This work is financially supported by the Nederlandse Organisatie voor Wetenschappelijk Onderzoek (NWO) as well as JSPS KAKENHI (Grant No. 19H006450). Y.-X. Zhang is supported by the European Unions Horizon 2020 research and
innovation program (Grant No. 712721, NanOQTech). We thank Xiang Zhang for useful discussions.
\end{acknowledgments}


\begin{thebibliography}{67}%
\makeatletter
\providecommand \@ifxundefined [1]{%
 \@ifx{#1\undefined}
}%
\providecommand \@ifnum [1]{%
 \ifnum #1\expandafter \@firstoftwo
 \else \expandafter \@secondoftwo
 \fi
}%
\providecommand \@ifx [1]{%
 \ifx #1\expandafter \@firstoftwo
 \else \expandafter \@secondoftwo
 \fi
}%
\providecommand \natexlab [1]{#1}%
\providecommand \enquote  [1]{``#1''}%
\providecommand \bibnamefont  [1]{#1}%
\providecommand \bibfnamefont [1]{#1}%
\providecommand \citenamefont [1]{#1}%
\providecommand \href@noop [0]{\@secondoftwo}%
\providecommand \href [0]{\begingroup \@sanitize@url \@href}%
\providecommand \@href[1]{\@@startlink{#1}\@@href}%
\providecommand \@@href[1]{\endgroup#1\@@endlink}%
\providecommand \@sanitize@url [0]{\catcode `\\12\catcode `\$12\catcode
  `\&12\catcode `\#12\catcode `\^12\catcode `\_12\catcode `\%12\relax}%
\providecommand \@@startlink[1]{}%
\providecommand \@@endlink[0]{}%
\providecommand \url  [0]{\begingroup\@sanitize@url \@url }%
\providecommand \@url [1]{\endgroup\@href {#1}{\urlprefix }}%
\providecommand \urlprefix  [0]{URL }%
\providecommand \Eprint [0]{\href }%
\providecommand \doibase [0]{http://dx.doi.org/}%
\providecommand \selectlanguage [0]{\@gobble}%
\providecommand \bibinfo  [0]{\@secondoftwo}%
\providecommand \bibfield  [0]{\@secondoftwo}%
\providecommand \translation [1]{[#1]}%
\providecommand \BibitemOpen [0]{}%
\providecommand \bibitemStop [0]{}%
\providecommand \bibitemNoStop [0]{.\EOS\space}%
\providecommand \EOS [0]{\spacefactor3000\relax}%
\providecommand \BibitemShut  [1]{\csname bibitem#1\endcsname}%
\let\auto@bib@innerbib\@empty
\bibitem {quantum_magnetism}A. Auerbach, \textit{Interacting Electrons and
Quantum Magnetism} (Springer-Verlag, New York, 1994).

\bibitem {RKKY}M. A. Ruderman and C. Kittel, Phys. Rev. \textbf{96}, 99
(1954); T. Kasuya, Prog. Theor. Phys. \textbf{16}, 45 (1956); K. Yosida, Phys.
Rev. \textbf{106}, 893 (1957).



\bibitem {Suhl_Nakamura1}H. Suhl, Phys. Rev. \textbf{109}, 606 (1958).

\bibitem {Suhl_Nakamura2}T. Nakamura, Progr. Theoret. Phys. (Kyoto)
\textbf{20}, 542 (1958).

\bibitem {optical_coherence}L. Mandel and E. Wolf, \textit{Optical Coherence
and Quantum Optics} (Cambridge University Press, Cambridge, England, 1995).



\bibitem {first_2010}\"O. O. Soykal, and M. E. Flatt\'e, Phy. Rev. Lett.
\textbf{104}, 077202 (2010).

\bibitem {second}H. Huebl, C. W. Zollitsch, J. Lotze, F. Hocke, M.
Greifenstein, A. Marx, R. Gross, and S. T. B. Goennenwein, Phy. Rev. Lett.
\textbf{111}, 127003 (2013).

\bibitem {third}Y. Tabuchi, S. Ishino, T. Ishikawa, R. Yamazaki, K. Usami, and
Y. Nakamura, Phy. Rev. Lett. \textbf{113}, 083603 (2014).

\bibitem {fourth}X. Zhang, C.-L. Zou, L. Jiang, and H. X. Tang, Phy. Rev.
Lett. \textbf{113}, 156401 (2014).

\bibitem {MagnonDarkModes}X. Zhang, C.-L. Zou, N. Zhu, F. Marquardt, L. Jiang,
and H. X. Tang, Nat. Comm. \textbf{6}, 8914 (2015).

\bibitem {Lambert_2016}N. J. Lambert, J. A. Haigh, S. Langenfeld, A. C.
Doherty, and A. J. Ferguson, Phys. Rev. A \textbf{93}, 021803(R) (2016).

\bibitem {Babak}B. Z. Rameshti and G. E. W. Bauer, Phys. Rev. B \textbf{97},
014419 (2018).



\bibitem {non_hermitian0}N. Hatano and D. R. Nelson, Phys. Rev. Lett.
\textbf{77}, 570 (1996).

\bibitem {non_hermitian2}C. M. Bender, D. C. Brody, and H. F. Jones, Phys.
Rev. Lett. \textbf{89}, 270401 (2002).

\bibitem {non_hermitian4}C. M. Bender, D. C. Brody, H. F. Jones, and B. K.
Meister, Phys. Rev. Lett. \textbf{98}, 040403 (2007).

\bibitem {non_hermitian3}D. Zhang, X.-Q. Luo, Y.-P. Wang, T.-F. Li, and J. Q.
You, Nat. Commun. \textbf{8}, 1368 (2017).

\bibitem {non_hermitian1}R. El-Ganainy, K. G. Makris, M. Khajavikhan, Z. H.
Musslimani, S. Rotter, and D. N. Christodoulides, Nat. Phys. \textbf{14}, 11 (2018).

\bibitem {non_hermitian5}S. Y. Yao and Z. Wang, Phys. Rev. Lett. \textbf{121},
086803 (2018).

\bibitem {non_hermitian6}Z. Gong, Y. Ashida, K. Kawabata, K. Takasan, S.
Higashikawa, and M. Ueda, Phys. Rev. X. \textbf{8}, 031079 (2018).

\bibitem {non_hermitian7}X. S. Yang, Y. Cao, and Y. Zhai, arXiv:1904.02492.

\bibitem {non_hermitian8}S. Y. Yao, F. Song and Z. Wang, Phys. Rev. Lett.
\textbf{121}, 136802 (2018).

\bibitem {non_hermitian9}H. Jiang, L. J. Lang, C. Yang, S. L. Zhu, and S.
Chen, Phys. Rev. B {\bf 100}, 054301 (2019).

\bibitem {Book}N. Moiseyev, \textit{Non-Hermitian Quantum Mechanics},
(Cambridge University Press, Cambridge, 2011).



\bibitem {level_attraction_Canming}M. Harder, Y. Yang, B. M. Yao, C. H. Yu, J.
W. Rao, Y. S. Gui, R. L. Stamps, and C.-M. Hu, Phys. Rev. Lett. \textbf{121},
137203 (2018).

\bibitem {level_attraction_Yu}B. M. Yao, T. Yu, X. Zhang, W. Lu, Y. S. Gui,
C.-M. Hu, and Y. M. Blanter, arXiv:1906.12142.



\bibitem {subradiance1}A. Asenjo-Garc\'ia, M. Moreno-Cardoner, A. Albrecht, H.
J. Kimble, and D. E. Chang, Phys. Rev. X \textbf{7}, 031024 (2017).

\bibitem {subradiance2}Y.-X. Zhang and K. M\o {}lmer, Phys. Rev. Lett.
\textbf{122}, 203605 (2019).

\bibitem {subradiance3}M. M. Cardoner, D. Plankensteiner, L. Ostermann, D. E.
Chang, and H. Ritsch, Phys. Rev. A {\bf 100}, 023806 (2019).

\bibitem {subradiance4}P.-O. Guimond, A. Grankin, D. V. Vasilyev, B.
Vermersch, and P. Zoller, Phys. Rev. Lett. \textbf{122}, 093601 (2019).

\bibitem {subradiance5}G. Buonaiuto, R. Jones, B. Olmos, and I. Lesanovsky, arXiv:1902.08525.

\bibitem {subradiance6}J. Li, S. Y. Zhu, and G. S. Agarwal, Phys. Rev. Lett.
\textbf{121}, 203601 (2018).

\bibitem {D1}L. M. Sieberer, S. D. Huber, E. Altman, and S. Diehl, Phys. Rev.
Lett. \textbf{110}, 195301 (2013).

\bibitem {D2}S. Diehl, A. Micheli, A. Kantian, B. Kraus, H. P. B\"uchler, and
P. Zoller, Nat. Phys. \textbf{4}, 878 (2008).

\bibitem {D3}N. Bernier, E. Torre, and E. Demler, Phys. Rev. Lett.
\textbf{113}, 065303 (2014).

\bibitem {Jackson}J. D. Jackson, \textit{Classical Electrodynamics}, (Wiley,
New York, 1998).

\bibitem {chiral_review}P. Lodahl, S. Mahmoodian, S. Stobbe, A.
Rauschenbeutel, P. Schneeweiss, J. Volz, H. Pichler, and P. Zoller, Nature
(London) \textbf{541}, 473 (2017).



\bibitem {chiral_optics1}F. Le Kien, S. D. Gupta, K. P. Nayak, and K. Hakuta,
Phys. Rev. A \textbf{72}, 063815 (2005).

\bibitem {chiral_optics5}B. le Feber, N. Rotenberg, and L. Kuipers, Nat.
Commun. \textbf{6}, 6695 (2015).

\bibitem {chiral_optics2}M. Scheucher, A. Hilico, E. Will, J. Volz, and A.
Rauschenbeutel, Science \textbf{354}, 1577 (2016).

\bibitem {chiral_optics4}B. Vermersch, P.-O. Guimond, H. Pichler, and P.
Zoller, Phys. Rev. Lett. \textbf{118}, 133601 (2017).



\bibitem {parametric_pumping1}A. G. Gurevich and G. A. Melkov,
\textit{Magnetization Oscillations and Waves} (CRC, New York, 1996).

\bibitem {parametric_pumping2}V. S. L'vov, \textit{Wave Turbulence under
Parametric Excitations: Applications to Magnetics} (Springer, Berlin, 1994).



\bibitem {Yu1}T. Yu, C. P. Liu, H. M. Yu, Y. M. Blanter, and G. E. W. Bauer,
Phys. Rev. B \textbf{99}, 134424 (2019).

\bibitem {Yu2}J. L. Chen, T. Yu, C. P. Liu, T. Liu, M. Madami, K. Shen, J. Y.
Zhang, S. Tu, M. S. Alam, K. Xia, M. Z. Wu, G. Gubbiotti, Y. M. Blanter, G. E.
W. Bauer, and H. M. Yu, arXiv:1903.00638.



\bibitem {non_local}Y. Tserkovnyak, A. Brataas, G. E. W. Bauer, and B. I.
Halperin, Rev. Mod. Phys., \textbf{77}, 1375 (2005).



\bibitem {Kamra_phonon}A. Kamra, H. Keshtgar, P. Yan, and G. E. W. Bauer,
Phys. Rev. B \textbf{91}, 104409 (2015).

\bibitem {Simon_phonon}S. Streib, H. Keshtgar, and G. E. W. Bauer, Phys. Rev.
Lett. \textbf{121}, 027202 (2018).

\bibitem {PRB}T. Yu, X. Zhang, S. Sharma, Y. M. Blanter, and G. E. W. Bauer,
submitted to Physical Review B.











\bibitem {Fano}U. Fano, Phys. Rev. \textbf{124}, 1866 (1961).

\bibitem {Mahan}G. D. Mahan, \textit{Many Particle Physics} (Plenum, New York, 1990).



\bibitem {Magnon_radiation}B. M. Yao, T. Yu, Y. S. Gui, J. W. Rao, Y. T. Zhao,
W. Lu, and C.-M. Hu, arXiv: 1902.06795.



\bibitem {input_output1}C. W. Gardiner and M. J. Collett, Phys. Rev. A
\textbf{31}, 3761 (1985).

\bibitem {input_output2}A. A. Clerk, M. H. Devoret, S. M. Girvin, F.
Marquardt, and R. J. Schoelkopf, Rev. Mod. Phys. \textbf{82}, 1155 (2010).



\bibitem {magnonics4}V. E. Demidov, S. Urazhdin, G. de Loubens, O. Klein, V.
Cros, A. Anane, and S. O. Demokritov, Phys. Rep. \textbf{673}, 1 (2017).
\end{thebibliography}
\end{document}